# Broadband impulsive stimulated Raman spectroscopy reveals electronic state-specific vibronic coupling and vibrational coherence transfer through nonadiabatic electronic coupling

Ramandeep Kaur[1], Shaina Dhamija[1], Garima Bhutani[1,2], Amit Kumar[1], Arijit K. De[1]*

[1]*Condensed Phase Dynamics Group*, Department of Chemical Sciences, Indian Institute of Science Education and Research Mohali, Knowledge City, Sector 81, SAS Nagar, Punjab, 140306.

[2]*Molecular Spectroscopy Laboratory*, RIKEN, 2-1 Hirosawa, Wako, Saitama 351-0198, Japan.

*Email: akde@iisermohali.ac.in

**Abstract:** Vibrational wavepacket dynamics in the ground ($X0_g^+$) and excited ($B0_u^+$) electronic states of iodine (in carbon tetrachloride) under impulsive-pump/broadband-probe excitation are revisited. A method for accurate chirp correction, necessary to determine the zero time for each component of spectrally dispersed data and thereby separate coherent vibrational dynamics from coherent artifacts and population kinetics, is introduced. While from these processed time-domain data the absolute Raman cross-section in the ground electronic state can be calculated using steady-state absorption, we show that the same can be done using the pump-probe data itself, and further extend this method as a benchmark to calculate the same for the excited electronic state; these cross-sections report on vibronic couplings specific to these states. Further, since the Fourier transform of the processed data yields information on vibrational modes averaged over the dephasing time, a wavelet analysis is performed to yield a joint time-frequency distribution of the vibrational modes, demonstrating how the time evolution of their frequencies (i.e., mode dispersion) can be extracted. The vibrational modes of the ground and excited electronic states are shown to exhibit distinct dispersion characteristics. Since overlapping spectral features appear at different time windows, such an analysis can disentangle spectral congestion, even from a simple one-dimensional measurement (i.e., only one parameter, the pump-probe delay, is varied). Most interestingly, a rapid time-dependent spectral shift and decay of the $B0_u^+$-state mode, followed by the appearance and growth of the $A'2_u$-state mode, directly correlates with the pre-dissociation, followed by 'solvent caging'-induced recombination. While such effects were reported from the ultrafast population kinetics in pump-probe experiments, the present work probes the dynamics by tracking the temporal evolution of vibrations in these states, and thus reveals transfer of vibrational coherence from one electronic state ($B0_u^+$) to another ($A'2_u$), mediated via nonadiabatic coupling to the intermediate dissociative state ($a1_g$), underscoring the importance of electronic coherence. Thus, our approach of 'visualizing' evolution of vibronic coherence from simple pump-probe measurements paves the way to explore such phenomena within complex molecular systems and materials.

## I. Introduction:

Deciphering chemical reactivity requires a detailed analysis of the concerted dynamics of electrons and nuclei [1–3]. In a typical pump-probe experiment, an ultrashort pump pulse creates coherent superpositions between vibrational levels in the ground and excited electronic states. Real-time monitoring of such non-stationary superpositions, or vibrational wavepackets, using a time-delayed continuum probe pulse has become routine [4–12]. Since the Fourier transform of these temporal oscillations yields the amplitudes and phases of vibrational modes, this approach of recording a Raman spectrum led to the development of time-domain Raman spectroscopy [13–15], known as impulsive stimulated Raman spectroscopy [16–25].

While femtosecond vibrational wavepacket dynamics have been studied for more than four decades, the implications of using a broadband probe in the patterns observed in the Raman spectra have been re-explored in recent years [26–38]. If the initial state prior to photo-excitation corresponds to the ground vibrational state (of the ground electronic state), the Stokes and coherent anti-Stokes pathways make the major contributions to the observed signal [39–41]. Following impulsive excitation by a pump pulse, Raman signals corresponding to the Stokes and coherent anti-Stokes pathways emerge on the red and blue sides of the probe transition frequency, respectively. Thus, for a continuum probe, the signals for any given mode appear over a broad spectral range of detection frequencies, as shown in **Figure 1**, and one would expect two situations: 1) If the probe has a spectral maximum or if there is an absorption maximum (corresponding to a resonant transition) or both, the Stokes and coherent anti-Stokes signals predominantly emerge at red and blue sides, respectively, and interfere at the maximum giving rise to a node on the spectra. 2) For a probe with a nearly flat spectral profile and in the absence of any resonant transition, these two signals appear everywhere and interfere constructively or destructively depending on their relative phase (which in turn depends on the chirp of the probe pulse), giving rise to a periodically modulated spectral intensity pattern (for a detailed discussion, see reference 35,37.

Despite this vast literature and recent explorations, several open questions remain in pump-probe measurements of vibrational coherence. For example, accurate chirp-correction, required to obtain correct spectral lineshapes, turns out to be rather difficult due to contributions from pump-scatter and 'coherent spikes' [42] near temporal overlap region of the pulses. While the vibrational coherence created in a specific (ground or excited) electronic state can be spectrally separated, to date, there is no report of any method to estimate electronic-state-specific absolute Raman cross sections to determine vibronic couplings. The most significant limitation of such third-order ($\chi^{(3)}$) technique is that since only one parameter (i.e., the delay between the pump and probe pulses) is varied, Fourier transform gives a time-averaged information of vibrational modes, but does not furnish any information on how the mode frequencies evolve (for which fifth-order ($\chi^{(5)}$) techniques are employed [43–46]).

In this work, we set out to address each of these prevailing issues. We resort to molecular iodine dissolved in carbon tetrachloride, a widely used system of choice because, although a diatomic molecule like iodine appears to be a rather simple system (since it has only one vibrational degree of freedom), the non-adiabatic coupling among its electronic surfaces, shown in **Figure 2**, renders it an ideal testbed for our experiment. We show how an analytic expression can accurately define the 'zero' of time. We also develop a novel approach to calculate absolute Raman cross sections pertaining to specific electronic states and, as a cautionary note, discuss why it is not correct to estimate the Huang-Rhys parameter from pump-probe data to quantify vibronic coupling. Above all, we introduce wavelet analysis to show how one can track the distinct time evolution of the frequencies of the ground- and excited-state modes, shedding light on the coherent transfer of vibrational wavepackets among coupled electronic states during predissociation, followed by recombination within the solvent cage [47]. We discuss how one can extend the methodologies developed in this work to uncover the role of vibronic coherence in problems involving more complex systems, such as excitonic energy transfer in photosynthesis [48].

## II. Experiment

The experimental setup uses a femtosecond transient absorption setup; details of which are reported elsewhere [35,37]. Briefly, the experimental setup uses a Ti: sapphire regenerative amplifier (Libra, Coherent Inc., ~800 nm, ~55 fs, 1 kHz) to pump a noncollinear optical parametric amplifier (Light Conversion, TOPAS White), generating 550 nm Raman pump pulses with a spectral full width half maximum (FWHM) of 423 cm⁻¹ (see **Figure S1**). A time-delayed broadband white-light, generated in CaF₂ (350–470 nm) or sapphire (470–740 nm), probes the sample, with polarization parallel to the pump, in a 1 mm cuvette. Pump (~2 mW) and probe (<10 µW) beams are spatially overlapped, and the self-heterodyned signal is recorded by blocking alternate pulses of the pump using a mechanical chopper operating at 500 Hz. The signal is spectrally dispersed and detected using a CMOS array detector.

## III. Results and Discussion

The spectra of the pump and probe and the absorption spectrum of I₂ in CCl₄ are shown in **Figure S1**. Raman pump centered at 550 nm resonantly excites iodine, populating the $B0_u^+$ electronic state in the vibrational levels $v' = 20 - 27$ (calculated using the Morse oscillator model and parameters given in **Table S3**). At the same time, it generates a wave-packet in the ground state $(X0_g^+)$, and non-resonantly excites the CCl₄ modes within its ground state.

### 1. Methodology

A detailed step-by-step procedure for obtaining mode dispersion is shown in **Figure 3**. The transient absorption contour map clearly shows the white-light probe's chirp, as shown in **Figure 3(a)**. Accurate chirp correction across the detection wavelength is necessary to determine the signal's initial phase, which in turn controls the spectral lineshape. In an earlier reported work [35,37], this was done by selecting the maxima or minima of the coherent spike as time-zero and discarding the data points within a few hundred femtoseconds. Since it is not a very accurate measure of time zero, we introduce a correction to the chirp with an analytic equation:

$$y(t) = y_0 + a_1 \times \cos\left(2\pi\nu(t - t_0) + \frac{\varphi\pi}{180}\right) \times e^{\frac{-(t-t_0)^2}{2\sigma^2}} + a_2 \times \frac{1}{2}\left(1 + erf\left(\frac{t-t_0}{\sqrt{2}\sigma}\right)\right) \quad (1)$$

Here, $t$ is the delay time, $y_0$ is the offset, $t_0$ is zero time, $\nu$ is the frequency, $\varphi$ is the phase, and $\sigma$ is the standard deviation; $a_1$ and $a_2$ are the contributions from the two parts. The equation consists of two parts: a single-frequency oscillatory function with a Gaussian envelope that accounts for coherent artifact, and a cumulative distribution function (i.e., an error function) describing the population rise. This equation is used to fit the initial part of the time-domain data to capture both the coherent artifact and the population rise, thereby obtaining an accurate time zero at a few detection wavelengths. A spectral slice of data along with this fit is shown in **Figure 3(a),** right panel, and the fitting parameters are tabulated in **Table S1**. Zero times at selected detection wavelengths are fitted using second-order polynomials in two probe spectral regions (measured with CaF$_2$ and sapphire). The resulting fits (parameters in **Table S2**) are used to interpolate the time-zero values across all other detection wavelengths (**Figure 3(a)**). The time-domain data are then chirp-corrected by finding the interpolated zero time at a particular detection wavelength on the original time axis and shifting the raw data accordingly **(Figure 3(b))**. Note that the oscillatory features are suppressed within the pump-scatter region (**Figures 3(a)** and **3(b)**). Therefore, in a few cases, the chirp-corrected data are first baseline-corrected by subtracting a background spectrum to remove pump scatter, thereby reviving the pump-probe signal.

The data obtained after chirp correction are truncated at the end to maintain a consistent data length for each detection wavelength. After skipping the initial data of 200 fs **(Figure 3(c))** (which is significantly contributed to by the coherent spike), the baseline is corrected using a tenth-order polynomial. Further, a Hanning window function is applied up to the last data point to ensure smooth damping of the data, which suppresses the side lobes around each spectral line (arising due to sudden truncation of oscillatory data), followed by zero padding to increase data points in the frequency domain **(Figure 3(d))**. These processed data are further analyzed using two different methods.

First, the processed data are Fourier-transformed (FT). Note that the choice of the order of the polynomial is set by the least contribution from the coherent spike (which gave the lowest zero frequency component after FT). The baseline correction using different orders of polynomials and a single exponential function, and the corresponding FT obtained, are shown in **Figures S2-S5** at selected detection wavelengths. The Fourier intensity map (i.e., contour of absolute square of the Fourier transform) of vibrational modes across detection wavelengths is shown in **Figure 3(e)** (left panel). To improve the signal-to-noise ratio, Fourier intensities are summed over the boxed region in **Figure 3(e).** The resulting frequency-domain spectrum is presented in **Figure 3(f)**, with a selected mode further highlighted in **Figure 3(g)**. This method yields very accurate frequency information, but loses all time information.

Second, continuous wavelet transform (CWT) analysis (**Figure 3(e),** right panel) is applied to the processed temporal data at all detection wavelengths. For each Raman mode identified in the FT contour (boxed regions in **Figure 3(e)** left panel), the corresponding CWT maps are summed over an appropriate range of detection wavelength to enhance the signal-to-noise ratio (**Figure 3(f)** right panel). This approach provides simultaneous time–frequency information. Representative spectral slices at selected delay times are shown in **Figure 3(g)**, right panel.

Throughout, coding, analysis, and plotting are done in MATLAB (R2024b).

## 2. Analysis of FT contour

Fourier Intensity maps in **Figure 3(e)** and **Figure 4** show numerous features grouped within the boxes and labeled I-X. Modes in the region I-IV, and VI are assigned to the excited state modes of iodine in the 100-130 cm$^{-1}$ region [49]. Region V (242 cm$^{-1}$) is assigned as the second harmonic of the 122 cm$^{-1}$ excited state mode. Region VII (212 cm$^{-1}$) is attributed to the ground state mode for iodine, where a node appears near the absorption maximum of iodine. The node appears because, under resonant excitation, the third field interaction comes from the frequency component of the broadband probe near the absorption maxima of the solute. The signals emerging in the red and blue-shifted frequencies are anti-phased in nature and correspond to Stokes and coherent anti-Stokes pathways, respectively. Region VIII (463 cm$^{-1}$) corresponds to the symmetric stretching mode of CCl$_4$, IX (218 cm$^{-1}$) and X (316 cm$^{-1}$) correspond to degenerate deformation modes of CCl$_4$. The sum of Fourier intensities over relevant regions of detection wavelength are shown **Figure S6**. We observe periodic intensity variation in the solvent modes because, under nonresonant excitation (for solvent), the third field interaction comes from many equally contributing frequencies due to the flat spectral profile of the broadband probe. The chirp of the probe introduces relative phase shifts among these frequency components: at certain detection wavelengths, red and blue-shifted contributions interfere destructively, while at others they interfere constructively. This detection wavelength-dependent interference leads to the observed periodic modulation of solvent mode intensity. A detailed discussion of these features is provided elsewhere [35].

## 3. Analysis of time-domain data

### 3.1 Stokes and coherent anti-Stokes signals for a specific mode

It is well known that the amplitude of oscillations ($A_{osc}$) in the time-domain data is directly proportional to the slope (i.e., the first derivative) of the absorption spectrum $\left(\frac{dOD}{d\omega}\right)$ [50,51]. The maximum intensity for a given mode is observed at detection wavelengths where the slope of the absorption spectrum is maximum or minimum, while a node appears at the absorption maximum. The absorption spectrum is first converted from the wavelength scale to the frequency scale and fitted with a lognormal function, as shown in **Figure 5(a)**. **Figure 5(b)** presents the intensity map of the ground-state iodine mode (212 cm$^{-1}$) as a function of detection wavelength, exhibiting a node at the absorption maximum and enhanced signal away from it. The derivative of the lognormal fit, along with the mode intensity as a function of detection frequency, is shown in **Figure 5(c)**. Obviously, the Raman spectral maxima nearly coincide with the maxima or minima of the derivative. The minimum slope corresponds to the coherent anti-Stokes contribution, while the maximum slope corresponds to the Stokes signal for the ground-state iodine mode. Note that at these two detection frequencies (3.39 × 10$^{15}$ s$^{-1}$ ~ 556 nm and 3.9 × 10$^{15}$ s$^{-1}$ ~483 nm), where the 212 cm$^{-1}$ mode intensities maximize, some modes other than this mode of interest also contribute. Therefore, the cumulative oscillations contributed by multiple modes at these detection frequencies do not exhibit a clear anti-phase relationship between the Stokes and coherent anti-Stokes pathways. To identify all these contributing modes, we first checked the FT at these detection frequencies (**Figure 5(d)**, then fitted the time-domain oscillations to a sum of cosine functions multiplied by an exponential function (attributed to dephasing of the oscillatory signal)

$$f(t) = a_1 \times cos\left(2\pi \times 6.36 \times t + \frac{\varphi_1 \pi}{180}\right) \times e^{-\frac{t}{\tau_1}} + a_2 \times cos\left(2\pi \times 13.85 \times t + \frac{\varphi_2 \pi}{180}\right) \times e^{-\frac{t}{\tau_2}} \quad (2)$$

with fixed frequencies corresponding to modes appearing at these detection frequencies (212 cm$^{-1}$ ~ 6.36 ps$^{-1}$ and 462 cm$^{-1}$ ~ 13.85 ps$^{-1}$), (**Figures 5(e) and 5(f)**, the fit parameters are shown in **Table S4)**. It can be seen from the oscillations extracted for the 212 cm$^{-1}$ at these detection frequencies show an anti-phased relationship, while the 463 cm$^{-1}$ mode does not show any such phase relationship (shown in **Figure 5(g) and (h)**).

While the above is an excellent method for analyzing the ground-state mode, if one tries to use the same for the excited state, there is no reference spectrum to start with. Therefore, we revisit the ground-state mode analysis using the ground-state bleach (GSB) signal of the pump-probe data. Since the GSB signal overlaps with the pump-scatter, the derivative of the spectra does not match the trend shown by the intensity of the ground-state mode. Therefore, the chirp-corrected data are first baseline-corrected by subtracting a background spectrum to remove pump scatter. The GSB spectrum at a selected time slice is then analyzed using the same procedure (**Figure 6**).

We obtain results similar to those from the absorption spectrum; however, some mismatch is observed between the detection frequency corresponding to the maximum intensity of the 212 cm$^{-1}$ mode and that predicted from the derivative of the optical density of the absorption spectrum (**Figure 5(c)**). In contrast, for the GSB signal, the agreement is exact (**Figure 6(c)**). Therefore, we use the pump-probe data itself for further analysis. This mismatch is because the absorption spectrum and the GSB spectrum are not fully identical. In absorption spectrum the excitation intensity is equal at all the wavelengths whereas in pump probe spectroscopy a narrow frequency bandwidth pump pulse is used and the shape of probe pulse does not affect the signal since it's effect in cancelled while taking differential OD, Thus the third field interaction is similar to the absorption spectrum, and the overall GSB signal is the combination of pump spectrum and the absorption spectrum.

In the ESA region, the strongest mode intensity appears on the coherent anti-Stokes side of the ESA spectrum, and no node is observed (**Figure 7**).

## 3.2 Quantifying electronic state-specific vibronic coupling

The Raman cross-section, particularly under resonant conditions, provides a direct probe of state-specific vibronic coupling, as our ISRS experiment interrogates the vibrational coherence generated in either the ground or excited electronic state. We calculate the differential Raman cross-section $\left(\frac{\partial \sigma}{\partial \omega}\right)$ for the ground-state mode of I$_2$ (212 cm$^{-1}$), at probe wavelengths $\omega$ corresponding to the maximum intensity of the coherent anti-Stokes signal, as well as on the Stokes side of the node associated with the absorption maximum of I$_2$ in CCl$_4$, using the equation:

$$\left(\frac{\partial^2 \sigma}{\partial \Omega \, \partial \omega}\right) = \ln 10 \cdot \frac{\hbar \omega_p}{E_p L N} \cdot \frac{n^2 \omega^2}{16\pi^3 c^2} \cdot \frac{1}{f_{xy}} \cdot A(\omega) \qquad (3)$$

where $\omega_p$ is the Raman pump frequency, $E_p$ is the energy of the incident Raman pump pulse, $L$ is the path length of the sample, $N$ is the number density of the liquid, $n$ is the index of refraction, $f_{xy}$ represents the spatial overlap of the pump and probe beams passing through the sample and $A(\omega)$ is the amplitude of Raman mode [52,53]. Here, we assume that the coherent signal is measured in the phase-matched direction, where the scattered intensity is maximized. Under this condition, the signal is effectively integrated over the solid angle; therefore, the measured quantity that corresponds to $\left(\frac{\partial \sigma}{\partial \omega}\right)$. Although there is some overlap between the Stokes and coherent anti-Stokes contributions on both sides, we assume that the intensity at the maximum slope is predominantly due to the Stokes pathway, while the intensity at the minimum slope is predominantly due to the coherent anti-Stokes pathway. Since the processed data time axis starts from 200 fs (to skip coherent artifact), the early time data is necessary to get a correct estimation of the Raman cross section. Therefore, we performed the FT of the fit to the time-domain oscillations, extrapolated to time zero (**Figure 6(e)** and **6(f)**), and then took the absolute square of the FT to obtain the spectral intensity of the 212 cm$^{-1}$ mode at two probe (detection) wavelengths as shown in **Figure S7**, which is then used to calculate the differential Raman cross-section. The differential Raman cross-section is proportional to the Fourier amplitude $A(\omega)$. The values obtained are $3.08 \times 10^{-28}$ m$^2$molecule$^{-1}$ at the probe wavelength corresponding to the coherent anti-Stokes maximum and $5.47 \times 10^{-28}$ m$^2$molecule$^{-1}$ at the Stokes maximum, as shown in **Section S3.1**. These values are approximately three orders of magnitude higher than those reported in the resonance Raman literature [54]. This is because resonance Raman involves resonance-enhanced spontaneous scattering via coupling to electronic transitions; the additional enhancement in ISRS reflects its coherent, stimulated nature, in which the transition rate is driven by the laser field intensity and reinforced by phase-matched, directional emission. We extend this analysis for the excited-state modes of I$_2$, and obtained differential Raman cross-section values of $2.53 \times 10^{-29}$ m$^2$molecule$^{-1}$.

### 3.3 Quantification of vibronic coupling through the Huang-Rhys parameter

The electron-phonon (or vibronic) coupling strength in a vibrating lattice is commonly described within the harmonic approximation using a displaced potential model. It can be quantified through three interrelated parameters: the equilibrium coordinate displacement ($Q_0^{(b)} - Q_0^{(a)}$), Huang-Rhys parameter ($S$), and reorganization energy ($E_{dis}$). The Huang Rhys parameter ($S$) is defined as

$$S = \frac{A^2}{2} = \frac{1}{2}\frac{M\omega^2}{\hbar\omega}(Q_0^{(b)} - Q_0^{(a)})^2 = \frac{E_{dis}}{\hbar\omega} \tag{4}$$

where $M$ is the effective mass, $\hbar$ is the reduced Planck constant, and $\omega$ is the vibrational frequency. The dimensionless parameter $A$ characterizes the difference in electron-lattice (here, vibronic) coupling between the two electronic states $a$ and $b$.

Although the Huang-Rhys parameter is traditionally defined for a crystalline lattice [55,56], here we use it for a diatomic molecule to quantify vibronic coupling within the harmonic approximation. We calculate $S$ for $X0_g^+ \rightarrow B0_u^+$ to estimate the vibronic coupling between the two states within the harmonic approximation using a displaced potential model. Here, we treat $\omega$ as the molecular normal mode frequency instead of the lattice frequency. In literature [57], $S$ has also been calculated from the amplitude of oscillation at time zero ($A_{osc}$) of time domain data at the detection wavelength corresponding to the maximum slope $\left(\frac{dOD}{d\omega}\right)$ of the steady-state absorption spectrum, where $A_{osc}$ is related to energy shift (frequency) $\Delta\omega$ by $A_{osc} = \left(\frac{dOD}{d\omega}\right)\Delta\omega$. The $S$ is calculated for the Stokes and coherent anti-Stokes pathways (at maximum and minimum slopes) using **equation 4,** and the calculations are shown in **Section S3.2**. The values of $S$ are 0.04 and 0.02 at the Stokes and coherent anti-Stokes maxima, respectively. Similarly, the same calculations are performed for ground-state bleach (GSB) and excited-state absorption bands of the pump-probe data of $I_2$/$CCl_4$. The chirp corrected pump-probe data (ISRS) is first baseline-corrected by subtracting the background spectra to remove pump scatter. Then, the GSB/ESA is converted to the frequency scale, fitted with a lognormal function at some delay time t, and the same procedure discussed above is repeated. S is calculated as 0.04 and 0.04 for the coherent anti-Stokes and Stokes sides of the node for the ground-state mode. For ESA, the value of 0.027 is obtained. The corresponding Figures, fit parameters, and calculations are shown in **Figures 6** and **7, Tables S5** and **S6, and Section S3.2.**

The Huang-Rhys parameter ($S$) is a measure of vibronic coupling (under the displaced-harmonic-oscillator approximation), which varies linearly with the Stokes shift [58,59] (the linear relationship breaks down at elevated temperature and beyond the Franck-Condon approximation [60]). Therefore, Stokes shift measurements are ideal to estimate this parameter. However, this vibronic coupling is assumed to be the same for each electronic state [55–60]. Since the GSB/ESA signal projects only the ground-/excited-state vibrational coherence, which have very different vibronic couplings, estimating the Huang-Rhys parameter from a single signal type in pump-probe measurements, a widely adopted practice in contemporary literature, is imprecise. Therefore, for pump-probe measurements, estimating the absolute Raman cross section is more appropriate.

### 3.4 Joint time-frequency analysis

The time-domain ISRS response does not provide frequency information directly (**Figure 3(a)**). Applying the Fourier transform yields accurate frequency information but results in a loss of temporal resolution (**Figure 3(e),** left panel). Consequently, neither approach alone can fully capture the time-dependent evolution of vibrational modes, motivating the use of time-frequency analysis methods, such as the continuous wavelet transform (CWT). CWT provides a combined time-frequency representation, albeit at the cost of reduced resolution in both domains (**Figures 3(f) and 3(g),** right panel). The resulting spectra at different delay times are baseline-corrected by subtracting a linear background from each spectrum and are subsequently fitted with a Voigt profile:

$$y = y_0 + (f_1 * f_2)(x) = y_0 + A \frac{2ln2}{\pi^{3/2}} \frac{w_L}{w_G^2} \int_{-\infty}^{\infty} \frac{e^{-t^2}}{\left(\sqrt{ln2}\frac{w_L}{w_G}\right)^2 + \left(\sqrt{4ln2}\frac{x-x_c}{w_G}-t\right)^2} dt \qquad (5)$$

Here, $y_0$ is the offset, $x_c$ is the center, $A$ is the area, $w_G$ is the Gaussian FWHM, $w_L$ is the Lorentzian FWHM. The time-dependent Raman mode peak position is extracted from the Voigt fit. In addition, the integrated area under the Voigt profile is obtained, and its temporal decay is fitted with a single- or bi-exponential function to extract the vibrational dephasing time of the mode.

**Figure 8(a)** shows the CWT map for ground state iodine mode (212 cm$^{-1}$), it seems that the spectrum is red shifting (**Figure 8(b)**) in the initial time, it could be due to the reason that initial spectra is dominated by high intensity mode (463 cm$^{-1}$) or some contributions coherent artifact are still present even after skipping the data for initial 200 fs. To avoid these effects, data points corresponding to an additional 250 fs are excluded from the peak-shift analysis and from the integrated-area fit to obtain more reliable estimates of the dephasing time. **Figure 8(c)** shows that this mode does not shift with time, and a biexponential fit of the integrated area yields time constants of 0.21 and 3.93 ps (**Figure 8(d)** and **Table S7**). Similarly, the solvent modes also do not show any time-dependent evolution of Raman frequencies, as shown in **Figures S8** and **S9**. This is because, in the ground electronic state of iodine (and solvent), only the first few vibrational levels, mainly the $v" = 0, 1, 2$ levels, are significantly populated (since $k_B T = 207 cm^{-1}$). The vibrational coherences being created at the bottom of the potential well, the anharmonic contribution is less, so the frequency difference between them, say, between $v" = 0$ & 1 and $v" = 1$ & 2, is not much. Therefore, the Raman frequency remains constant with time, and no dynamic frequency shift is observed.

A similar analysis is performed for excited-state modes as well. The FT map of the ESA region is divided into four regions, I, II, III, and IV, which represent distinct spectral features associated with vibrational wavepackets in different energy regions of the $B0_u^+$ state (shown in **Figure 4(a)** and zoomed in **Figure S10**. The probe couples the wavepacket created in different regions of $B0_u^+$ state to some high lying state such as $D'2_g$, $\beta 1_g$, $E0_g^+$ or $f0_g^+$ (**Figure S15**). The results of wavelet analysis for excited state modes are shown in **Figures 9, S11-S13,** and **Table S7**.

The excited-state vibrational modes of iodine exhibit clear mode dispersion, and their coherence lifetimes are significantly shorter than those of the ground state. In general, for the excited electronic state, the pump pulse creates coherence between any two vibrational levels other than the ground vibrational level. Since the anharmonicity constant for $B0_u^+$ state is higher than $X0_g^+$ state, and the pump pulse excites a large number of vibrational levels, i.e., $v' = 20 - 27$ levels, lying well above the bottom of $B0_u^+$ state, the anharmonic contributions are more significant compared to the ground-state, so the frequency differences among the vibrational coherences is large, leading to a dynamic frequency shift. The superposition of these coherences leads to a time-dependent evolution of the observed Raman frequency, which appears as a dynamic shift (mode dispersion) in the Raman frequencies. Since the coherence between upper vibrational levels dephases more quickly than between the lower ones, an overall blue shift is expected due to anharmonicity. This is observed for regions I, II, and III shift as shown in **Figures S11-S13**. At lower vibrational energy levels, the coherence lifetime increases, as evidenced in **Table S7**, indicating that the dephasing time becomes longer when the wavepacket evolves into and is probed in lower-energy regions.

Unlike the other regions, Region IV exhibits a unique behavior characterized by two distinct peaks: one that rapidly red-shifts and decays, followed by the emergence and growth of an unshifted feature at 119 cm$^{-1}$ beginning at ~1.4 ps, and another that remains relatively stationary in frequency while its intensity gradually diminishes. **(Figure 9 and S15)**.

Molecular iodine in condensed phase has been a topic of intense experimental and theoretical research over the decades [61,62]. Early picosecond studies reported that the $B0_u^+$ state of molecular iodine in the liquid phase undergoes predissociation with timescales shorter than 1 ps [47,63,64]. With the advent of femtosecond spectroscopy, transient dichroism and birefringence measurements revealed that vibrational coherences prepared in the $B0_u^+$ state decay with a characteristic time of ~230 fs, and that predissociation proceeds predominantly via coupling to the $a1_g$ surface [10]. These findings were further supported by fluorescence-detected femtosecond pump-probe spectroscopy on iodine in solid matrices and liquid phase and nonadiabatic molecular dynamics simulations [65–70]. As the iodine molecule dissociates, solvent cage dynamics play a crucial role in directing the outcomes, channeling the system to either escape (dissociation) or towards recombination into the excited state $A'2_u$ or the ground state $X0_g^+$. Moreover, because dipole coupling between the $A'2_u$ state and the $X0_g^+$ states is symmetry-forbidden, the population/coherence remains trapped in the $A'2_u$ state for extended durations [10].

Based on these reports, we have the following explanation, which is explained in **Figure 9(f)**. The two observed peaks arise from the bifurcation of the wavepacket due to nonadiabatic coupling between the $B0_u^+$ and $a1_g$, states. The unshifted component originates from the fraction of the wavepacket remaining in the $B0_u^+$ state, whereas the red-shifted component is associated with the fraction of the wavepacket that leaks into the $a1_g$ state via predissociation. The surrounding solvent molecules create a 'flexible cage' that opposes dissociation by forcing fragments to undergo geminate recombination into the excited states $A'2_u$ state. Since electric dipole coupling between the $A'2_u$ state and the $X0_g^+$ state is forbidden, this could account for the vibrational coherences remaining trapped in the $A'2_u$ state. Therefore, the appearance of an unshifted component at 119 cm$^{-1}$ at 1.4ps, along with its increased intensity following the decay of the red-shifted peak, is attributed to contributions from the $A'2_u$ electronic state. We also observe a red shift in the CWT (**Figure S16**) corresponding to region VI in the Fourier intensity map (**Figure 4(b)**), which we also attribute to the predissociation. Note that no significant initial frequency shift is observed in the excited-state modes. We attribute this to nuclear inertia, since the excitation to the $B0_u^+$ state is abrupt, the early-time motion of the wavepacket remains confined to the locally harmonic region of the potential.

We also observe distinct vibrational signatures, including overtones $2\omega$, $3\omega$, or combination bands in the wavelet analysis of the ESA region (**Figure S17**), arising from both solute and solvent modes; the solvent contributions are consistent with earlier reports [71]). These features are further corroborated by the Fourier intensity map (**region V, Figure 4(a)**), which shows the second harmonic of the ESA mode.

## IV. Further Discussion

The wavelet analysis introduced in the preceding section needs some critical discussion. First, information about the dynamic evolution of individual modes is obtained over a time window limited by the overall (homogeneous and inhomogeneous) dephasing time. Second, the different dispersive nature of the ground- and excited-state modes reinforces the argument that vibronic couplings differ across electronic states. Thus, the wavelet analysis could be applied to identify ground or excited-state modes of the same frequency based on their dispersion characteristics. Third, although only one parameter (i.e., the pump-probe delay) is varied, as in multidimensional spectroscopy, one can untwine overlapping spectral contributions because they appear at different time frames. Fourth, although the timescale of predissociation occurs on the vibrational period of the $B0_u^+$ state [10], the signal at a given time must arise from the undissociated molecules in the ensemble. Fifth, the significant differences between our approach and earlier pump-probe experiments [10,65,66,70] are that: 1) in earlier studies, signals were analyzed for specific probe wavelengths which are contributed by several vibrational modes whereas the broadband ISRS scheme allows to analyze the total contribution to the signal over the entire probe wavelengths for a specific mode at a time, and 2) we probed only bound states whereas since we analyzed specific vibrational modes contributed by only bound states, unlike the previous studies we didn't pick up any signal when the wavepacket is propagated in the continuum of the dissociative state (assuming there are no resonances). Therefore, our approach not only just complements previous pump-probe measurements of predissociation/caging dynamics, but it also provides an unambiguous mode-specific analysis of the classic problem.

Since the initial vibrational coherence was created only by the pump pulse (and the probe only interrogates this initially created coherence, the disappearance of $150\ cm^{-1}$ mode with concomitant its red-shift, followed by appearance of the $119\ cm^{-1}$ mode with no spectral shift, hints to vibrational coherence transfer between electronic states. One crucial observation is that due to the local nature of the nonadiabatic coupling [68], only part of the wavepacket leaks (and shows a red shift) while the other parts are preserved (and do not show any spectral shift), presumably because within a specific well (i.e., electronic state) vibrational coherence between upper levels do not get transferred to lower levels due to frequency mismatch owing to anharmonicity. Therefore, the only way the relaxation can happen is through the subtle interplay of electronic coherence between the participating states. We note that similar deliberations occurred in the context of photosynthetic energy transfer when long-lived coherent oscillations observed in pump-probe transients [72] were explored by multidimensional electronic spectroscopy and were attributed to electronic coherence [48,73], which were initially debated as primarily vibrational in nature [74], but later their vibronic nature was emphasized [75]. Subsequently, the transfer of vibronic coherence was also identified [76]. While in multidimensional spectroscopy, these couplings are embodied as off-diagonal cross-peaks whose amplitudes oscillate as a function of the waiting time (i.e., the pump-probe delay), a wavelet analysis approach similar to ours can still be applied to gain more insights into the nature of these coherent oscillations.

As a final note, the disappearance of vibrational coherence (from one sub-space) is comparable with the 'collapsing' of a wavefunction via 'decoherence', while its reappearance (in another sub-space) invokes the idea of its 'un-collapsing' [77]. Along this direction, one can think of interesting applications of our method in quantum computing and quantum information science [78].

## V. Conclusion

In summary, in the present work, we have introduced several new concepts: First, an accurate measure for chirp correction using an analytic expression is presented. Second, using the differential absorption from pump-probe data, electronic-state-specific absolute Raman cross-sections are measured. Third, the continuous wavelet transform is shown to disentangle overlapping spectral features, thereby revealing coherent vibrational coherence transfer between electronic states. Besides providing a detailed step-by-step method for applying each of these new concepts, we deliberated why the intensities of Raman modes in a given electronic state directly correlate with the vibronic coupling of that state, and why one should adopt this approach over the conventional approach of estimating the Huang-Rhys parameter from pump-probe data. Considering the power of joint time-frequency analysis, we envisage far-reaching applications of this method in exploring coherent energy/charge transfer within systems of varying complexities.

**Acknowledgements**

This work was funded by SERB (Grant No: CRG/2021/003981) and IISER Mohali (start-up grant). RK and AK thank UGC, SD thanks IISER Mohali, and GB thanks CSIR for the graduate fellowship.

[25] U. Megerle, I. Pugliesi, C. Schriever, C.F. Sailer, and E. Riedle, "Sub-50 fs broadband absorption spectroscopy with tunable excitation: putting the analysis of ultrafast molecular dynamics on solid ground," Appl. Phys. B **96**(2), 215–231 (2009).

[26] M. Liebel, C. Schnedermann, T. Wende, and P. Kukura, "Principles and Applications of Broadband Impulsive Vibrational Spectroscopy," J. Phys. Chem. A **119**(36), 9506–9517 (2015).

[27] J.C. Dean, S.R. Rather, D.G. Oblinsky, E. Cassette, C.C. Jumper, and G.D. Scholes, "Broadband Transient Absorption and Two-Dimensional Electronic Spectroscopy of Methylene Blue," J. Phys. Chem. A **119**(34), 9098–9108 (2015).

[28] J.A. Cina, P.A. Kovac, C.C. Jumper, J.C. Dean, and G.D. Scholes, "Ultrafast transient absorption revisited: Phase-flips, spectral fingers, and other dynamical features," J. Chem. Phys. **144**(17), 175102 (2016).

[29] S.R. Rather, and G.D. Scholes, "Slow Intramolecular Vibrational Relaxation Leads to Long-Lived Excited-State Wavepackets," J. Phys. Chem. A **120**(34), 6792–6799 (2016).

[30] C.C. Jumper, P.C. Arpin, D.B. Turner, S.D. McClure, S.R. Rather, J.C. Dean, J.A. Cina, P.A. Kovac, T. Mirkovic, and G.D. Scholes, "Broad-Band Pump–Probe Spectroscopy Quantifies Ultrafast Solvation Dynamics of Proteins and Molecules," J. Phys. Chem. Lett. **7**(22), 4722–4731 (2016).

[31] L. Monacelli, G. Batignani, G. Fumero, C. Ferrante, S. Mukamel, and T. Scopigno, "Manipulating Impulsive Stimulated Raman Spectroscopy with a Chirped Probe Pulse," J. Phys. Chem. Lett. **8**(5), 966–974 (2017).

[32] I. Gdor, T. Ghosh, O. Lioubashevski, and S. Ruhman, "Nonresonant Raman Effects on Femtosecond Pump–Probe with Chirped White Light: Challenges and Opportunities," J. Phys. Chem. Lett. **8**(8), 1920–1924 (2017).

[33] G. Batignani, C. Ferrante, G. Fumero, and T. Scopigno, "Broadband Impulsive Stimulated Raman Scattering Based on a Chirped Detection," J. Phys. Chem. Lett. **10**(24), 7789–7796 (2019).

[34] P.C. Arpin, and D.B. Turner, "Signatures of Vibrational and Electronic Quantum Beats in Femtosecond Coherence Spectra," J. Phys. Chem. A **125**(12), 2425–2435 (2021).

[35] S. Dhamija, G. Bhutani, A. Jayachandran, and A.K. De, "A Revisit on Impulsive Stimulated Raman Spectroscopy: Importance of Spectral Dispersion of Chirped Broadband Probe," J. Phys. Chem. A **126**(7), 1019–1032 (2022).

**Figures**

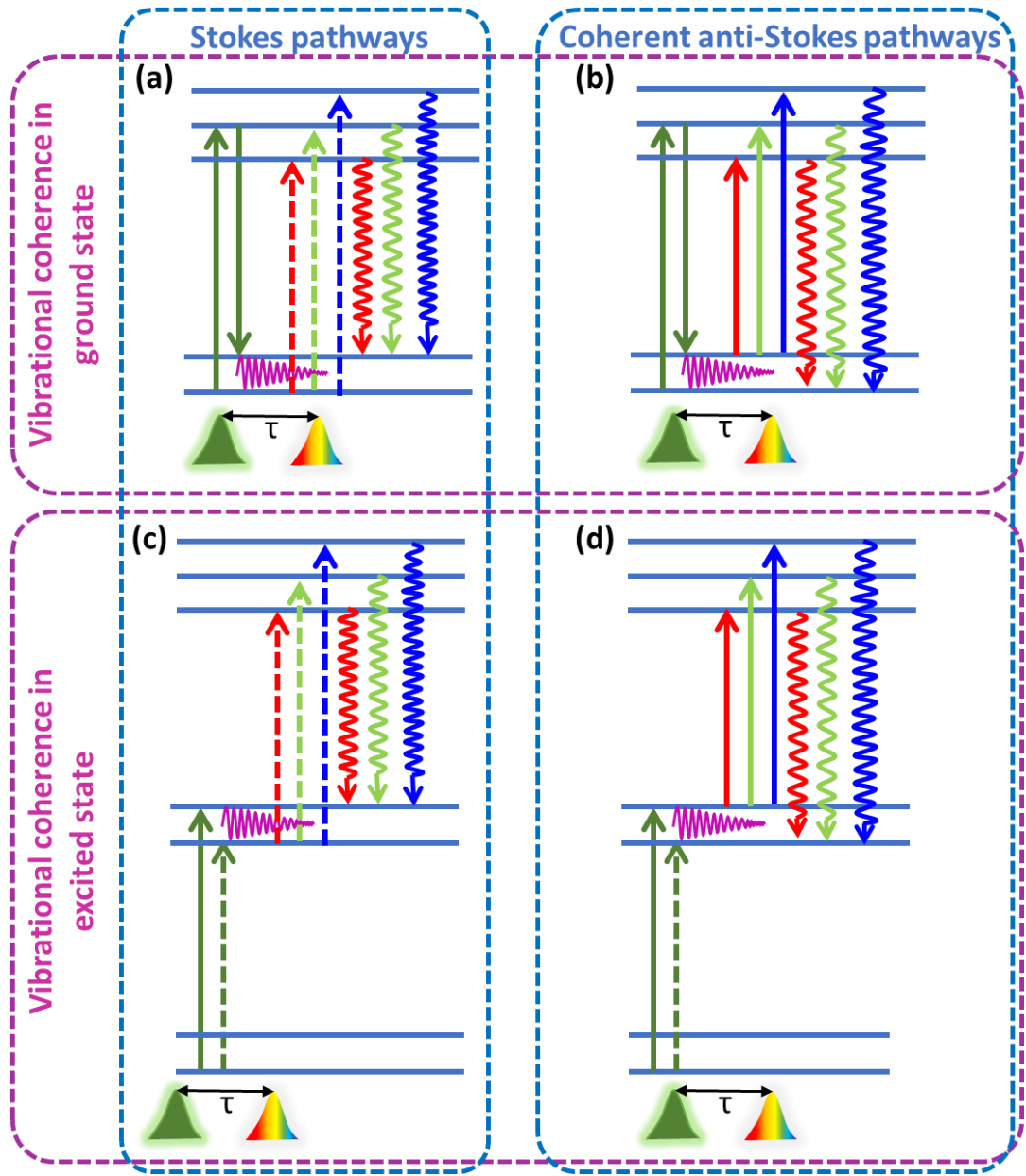

**Figure 1**: Arrow diagrams illustrating broadband impulsive stimulated Raman scattering. Pathways for the creation of vibrational coherence are shown for the ground state (a,b) and the excited state (c,d). Panels (a) and (c) correspond to Stokes pathways, while (b) and (d) represent coherent anti-Stokes pathways (with respect to the probe pulse), over a range of detection frequencies. Vertical solid/broken arrows denote interactions of the field with the ket (bra) side of the density matrix, and vertical wavy arrows indicate the emitted signal.

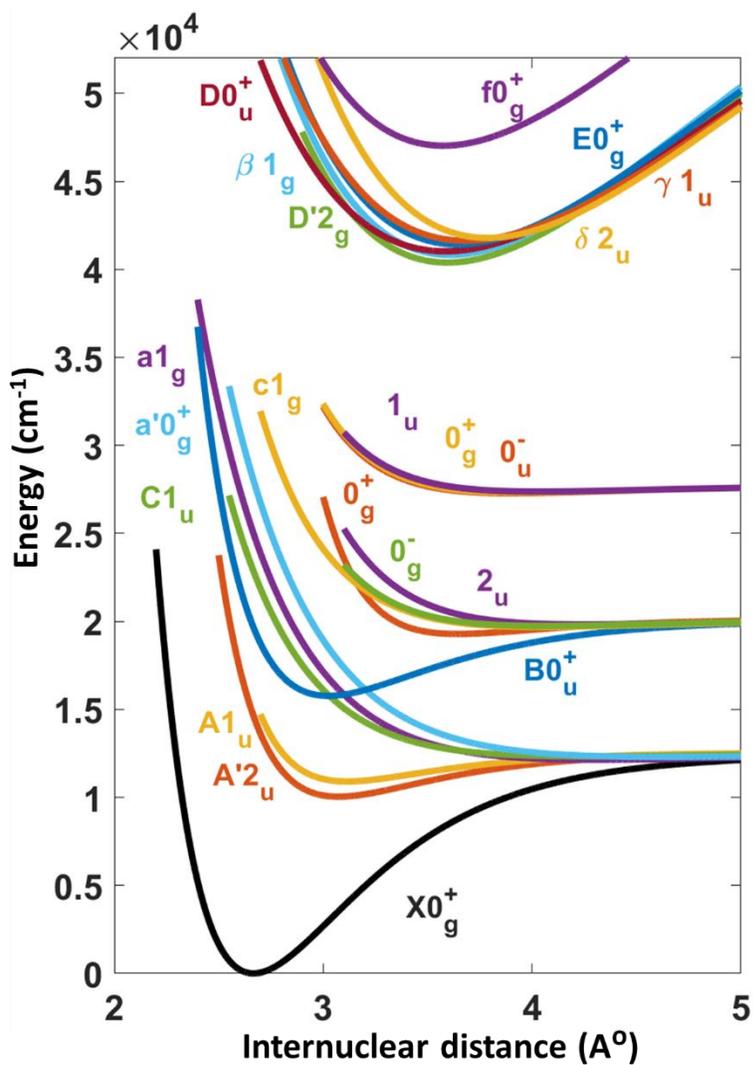

**Figure 2**: Schematic energy-level diagram of $I_2$ illustrating probe wavelengths accessing different regions (I–IV) of the $B0_u^+$ state; blue, green, yellow, and red arrows correspond to probe wavelengths of 355, 402, 425, and 460 nm, respectively. The parameters used to construct the diagram are taken from the literature, which are included in **Table S3**.

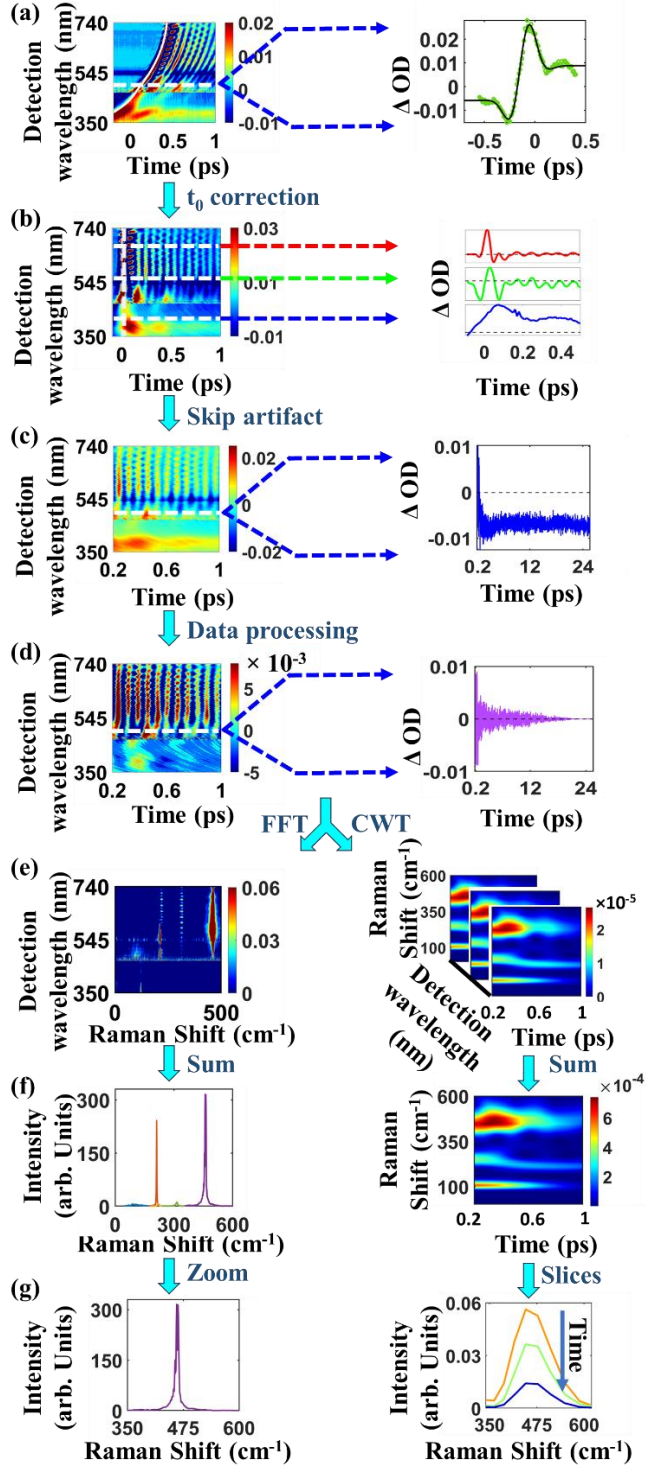

**Figure 3**: Schematic of the analysis workflow. (a) Raw time-domain contour map with $t_0$ interpolation (left); a spectral slice with eq 1 fit (right), (b) $t_0$-corrected data (left); slices at three detection wavelengths (right), (c) After coherent artifact removal (left), representative slice (right), (d) Free induction decay after baseline correction, Hanning windowing, and zero padding (left); corresponding slice (right), (e) Fourier intensity map (left); CWT at selected detection wavelengths (right), (f) summed signal over boxed region yielding no time information (left); Summed CWT slices (right) over appropriate detection wavelength region, (g) Zoomed Raman mode giving no time information s(left); slices at four delay times showing time–frequency information (symbols+lines, CWT; lines, Voigt fits) (right).

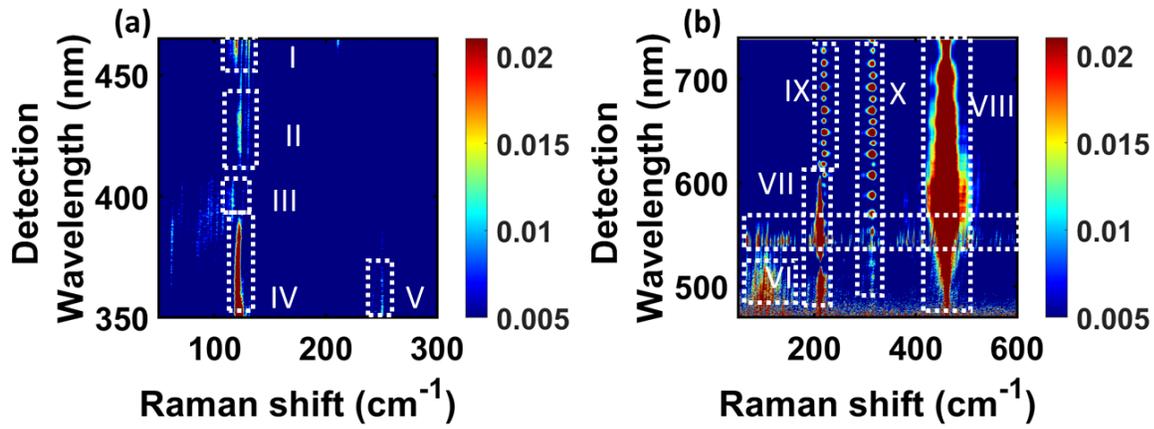

**Figure 4**: Fourier intensity map (a) at detection wavelengths from 350 nm to 470nm, (b) 470 nm to 735 nm; detection wavelength region 525nm to 735 nm corresponds to GSB band; 420-525 nm region has overlapping ESA and GSB bands, below 420 nm is the ESA band only, and yellow box represents the pump scatter. Similar spectral features are reported in the reference [37]; however, those studies presented amplitudes, whereas here the data are shown as intensity.

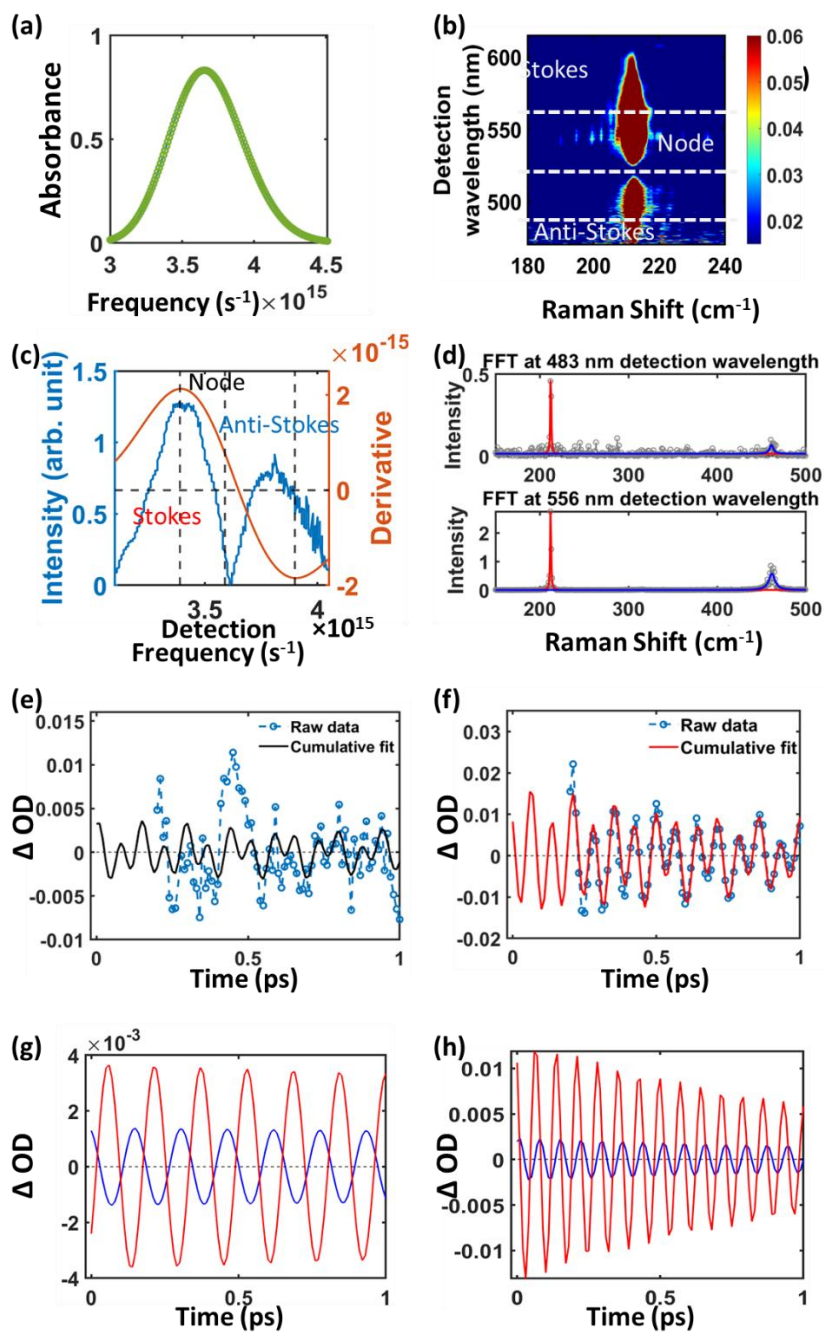

**Figure 5**: (a) Absorption spectrum of $I_2/CCl_4$ with lognormal fit, (b) Fourier intensity map corresponding to the ground-state iodine mode, (c) Derivative of lognormal fit of absorption spectrum (brown) and intensity of ground state iodine mode as a function of detection frequency (blue), (d) FT of time domain data at detection wavelength corresponding to maximum and minimum slope, (e), (f) Time-domain oscillations at selected detection wavelengths (483 and 556 nm), with scatter points representing raw data and solid lines showing fits to a sum of two cosine functions, each multiplied by an exponential function, (g) Oscillations of the 212 cm$^{-1}$ mode, showing anti-phase behavior, and (h) Oscillations of the 463 cm$^{-1}$ mode, exhibiting no clear phase relationship, extracted from temporal slices on either side of the node at 483 nm (blue) and 558 nm (red).

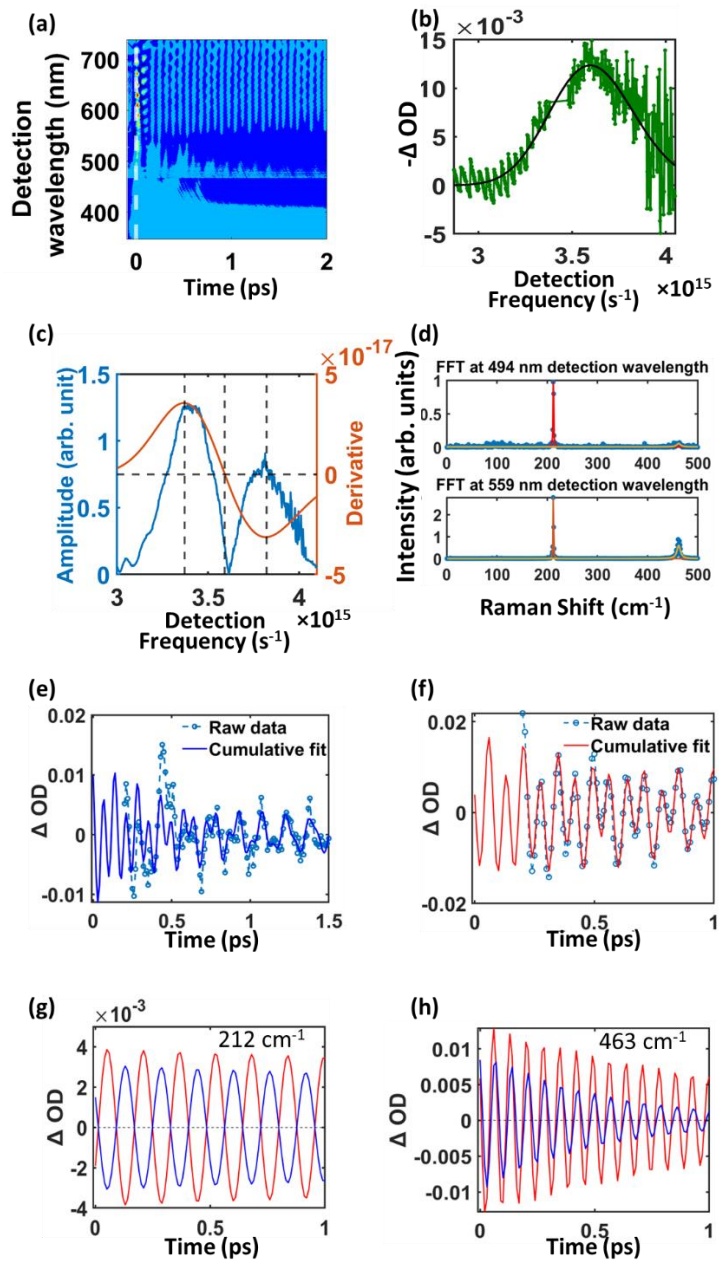

**Figure 6**: (a) Raw time domain ISRS data, (b) Ground state bleach spectrum (green) of $I_2/CCl_4$ with lognormal fit (black), (c) Derivative of lognormal fit of GSB signal (brown) and intensity of ground state iodine mode as a function of detection frequency (blue), (d) FT of time domain data at detection wavelength corresponding to maximum and minimum slope, (e), (f) Time-domain oscillations at selected detection wavelengths (494 and 559 nm), with scatter points representing raw data and solid lines showing fits to a sum of two cosine functions, each multiplied by an exponential function, (g) Oscillations of the 212 cm$^{-1}$ mode, showing anti-phase behavior, and (h) Oscillations of the 463 cm$^{-1}$ mode, exhibiting no clear phase relationship, extracted from temporal slices on either side of the node at 494 nm (blue) and 559 nm (red).

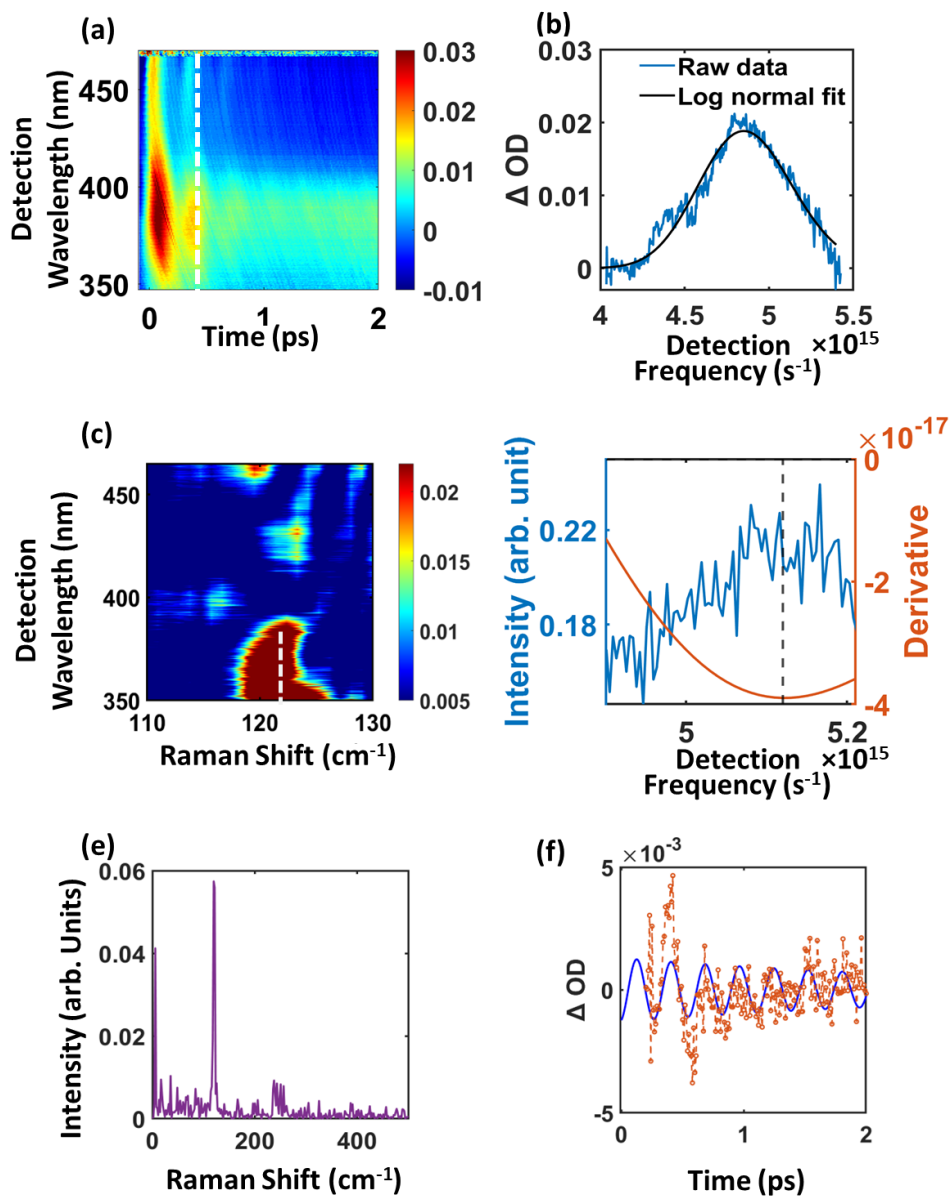

**Figure 7**: (a) Raw time domain ISRS data, (b) ESA spectrum (blue) of $I_2/CCl_4$ with lognormal fit (black), (c) ) FFT contour corresponding to ESA region, (d) derivative of ESA lognormal fit (red) and intensity of excited state iodine mode as a function of detection frequency (blue), (e) FT of the time-domain data at the detection wavelength corresponding to the maximum negative slope, (f) Time-domain oscillations at 368 nm: fit (blue) and raw data (brown).

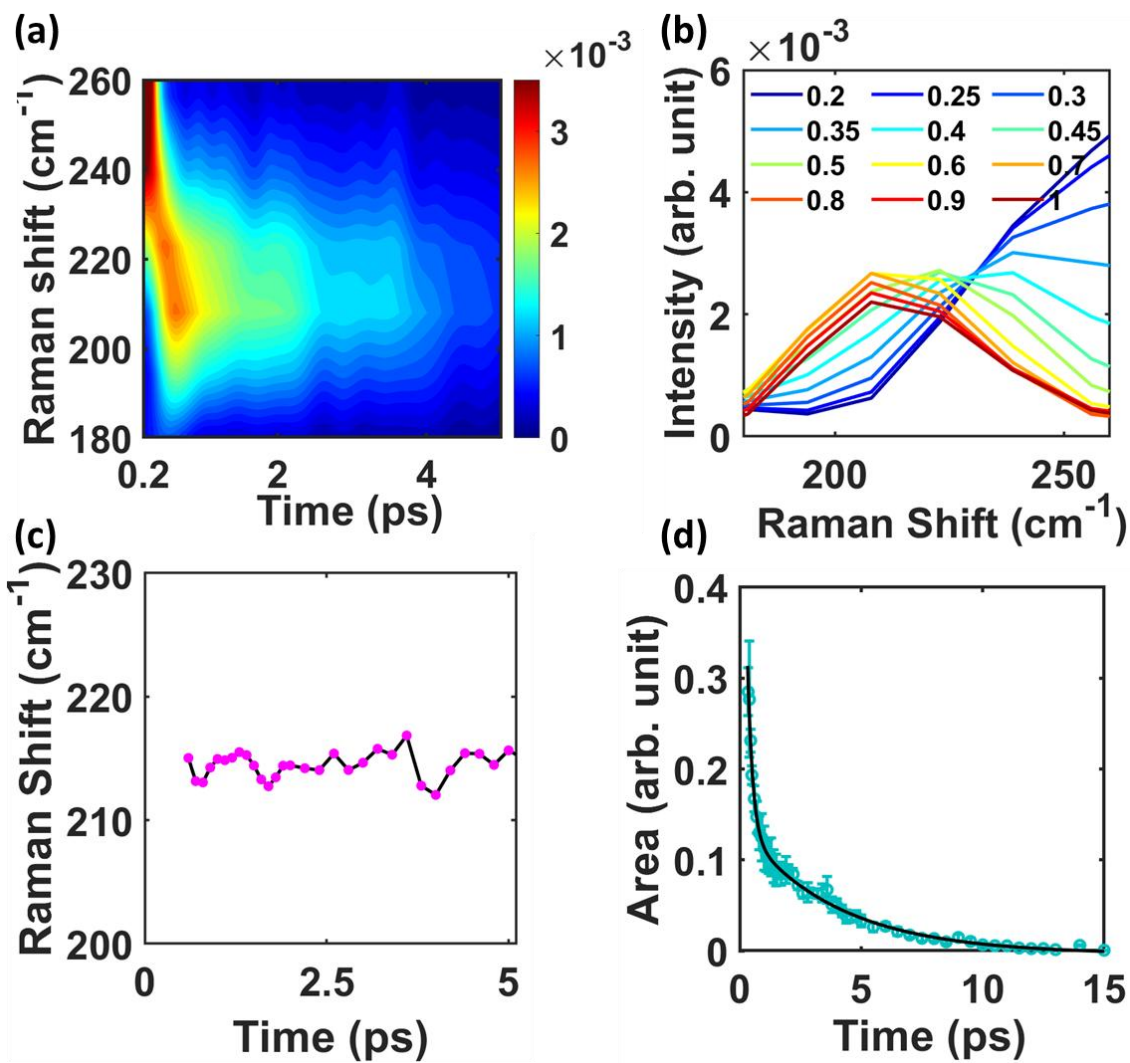

**Figure 8**: (a) CWT map for ground state iodine mode (212 cm$^{-1}$), (b) CWT slices at different delay times, (c) Raman peak shift as a function of time, (d) Decay of the integrated Voigt-fit area of the mode, reflecting its vibrational dephasing dynamics.

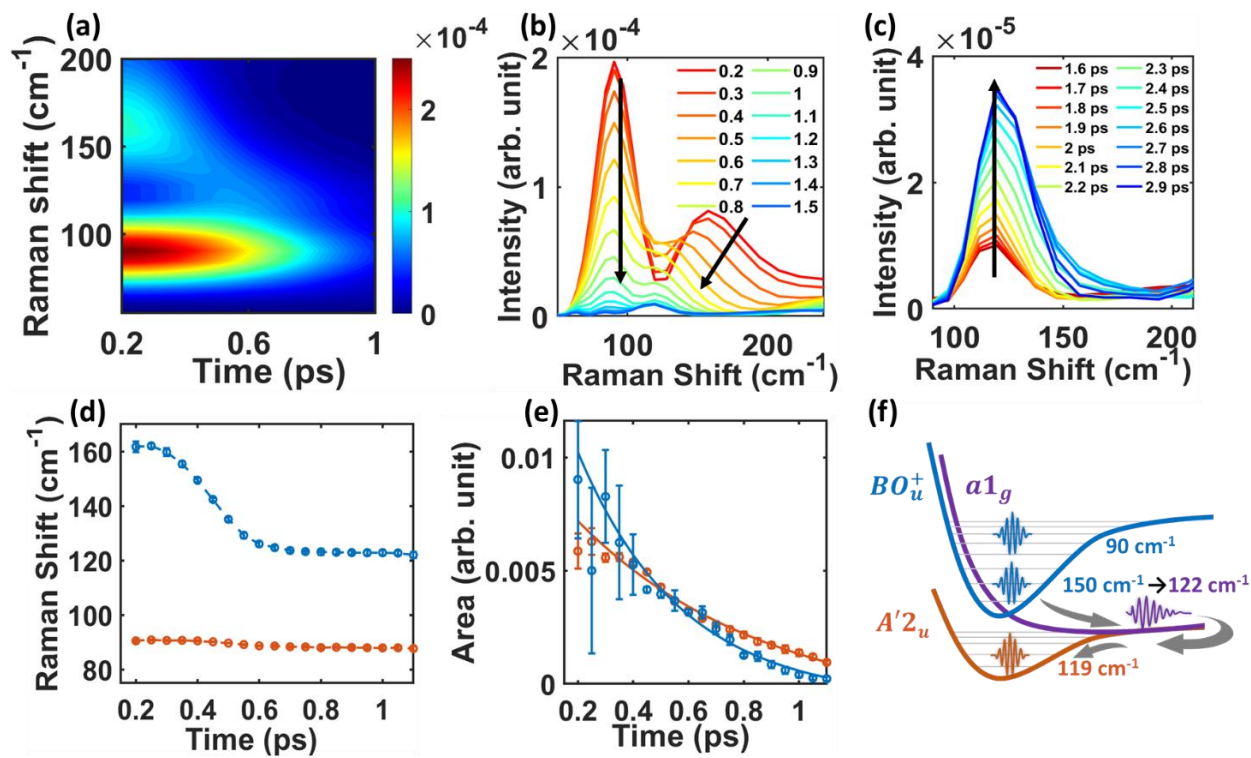

**Figure 9**: (a) CWT (summed over 350-390 nm detection wavelength) map for Region IV, (b) Slices of CWT map at different time delays, (c) revival of excited state mode at 119 cm$^{-1}$, (d) peak 1 (blue) and 2 (brown) shift as a function of time delays, (e) Decay of the integrated Voigt-fit area, reflecting its vibrational dephasing dynamics of the two modes, (f) Schematic illustration of predissociation via the a1$_g$ state, followed by transfer of vibronic coherence to the A′2$_u$ state due to the solvent cage effect.